\documentclass{article}
\pdfoutput=1

     \PassOptionsToPackage{numbers, compress}{natbib}

\usepackage[final]{icbinb_neurips_2020}

\usepackage[utf8]{inputenc} 
\usepackage[T1]{fontenc}    
\usepackage{hyperref}       
\usepackage{url}            
\usepackage{booktabs}       
\usepackage{amsfonts}       
\usepackage{nicefrac}       
\usepackage{microtype}      
\usepackage{comment}
\usepackage{amsmath}        
\usepackage{amsfonts}
\usepackage{enumitem}
\usepackage{tabularx}
\DeclareMathOperator{\sgn}{sgn}

\usepackage[pdftex]{graphicx}
\usepackage{subcaption}

\DeclareSymbolFont{bbold}{U}{bbold}{m}{n}
\DeclareSymbolFontAlphabet{\mathbbold}{bbold}

\usepackage{array}
\newsavebox{\bitbucket}
\newcolumntype{V}{>{\hspace{-2\tabcolsep}\savebox{\bitbucket}\bgroup}{c}<{\egroup}}

\title{I'm Sorry for Your Loss:\\
Spectrally-Based Audio Distances Are Bad at Pitch}

\author{%
  Joseph Turian\\
  \texttt{lastname@gmail.com} \\
  \And
  Max Henry \\
  \texttt{maxsolomonhenry@gmail.com} \\
}

\begin{document}

\maketitle

\begin{abstract}
Growing research demonstrates that synthetic failure modes imply poor generalization.
We compare commonly used audio-to-audio losses on a synthetic benchmark, measuring the pitch distance between two stationary sinusoids. The results are surprising: many have poor sense of pitch direction.
These shortcomings are exposed using simple rank assumptions. Our task is trivial for humans but difficult for these audio distances, suggesting significant progress can be made in self-supervised audio learning by improving current losses.
\end{abstract}

\section{Introduction}
\label{sec:introduction}

Rather than a physical quantity contained in the signal, pitch is a {\em percept} that is strongly correlated with the fundamental frequency of a sound. While physically elusive, pitch is a natural dimension to consider when comparing sounds.
Children as young as 8 months old are sensitive to melodic contour
\cite{trehub_infants_1984}, implying that an early and innate sense of relative pitch is important to parsing our auditory experience. This basic
ability is distinct from the acute pitch sensitivity of professional musicians, which improves
as a function of musical training \cite{kishon-rabin_pitch_2001}.

Pitch is important to both music and speech signals. 
Typically, speech pitch is more narrowly circumscribed than music, having a range of 70--1000Hz, compared to music which spans roughly the range of a grand piano, 30--4000Hz \cite{hess_pitch_2012}. Pitch in speech signals also fluctuates more rapidly \cite{carey_comparison_1999}.
In tonal languages, pitch has a direct influence on the  meaning of words \cite{47790,yasuda_investigation_2019, DBLP:conf/interspeech/0006LLWCY20}. For non-tone languages such as English, pitch is used to infer (supra-segmental) prosodic and speaker-dependent features \cite{frick_communicating_1985,DBLP:conf/icassp/GarbaceaOLLLVW19, DBLP:conf/interspeech/0006LLWCY20}.

In this work, we demonstrate pitch-based failure modes for audio-to-audio distances commonly used as loss functions. Negative results have potentially deleterious consequences on this benchmark: gradient-descent takes a circuitous route for some losses, implying that generalization speed and final accuracy are impaired. Worse yet, other loss functions get stuck in varying scopes of local minima. Their search spaces contain both fine-grained oscillations and broader plateaus.

\subsection{Perceptual Audio-to-Audio Distances}
\label{sec:audio-losses}

Audio-to-audio distances attempt to quantify what a human judge would consider the similarity between sounds. Therefore, we argue that they are fundamentally perceptual in nature. 
The notion of a distance between two sounds has a long history of investigation in perceptual research \cite{stevens_scale_1936, stevens_scale_1937}. While early work focused on loudness and pitch distances in simple tones, a more recent thread in experimental psychology explores the auditory distances between instrument timbres, sounds that are more abstractly connected \cite{grey_multidimensional_1977,mcadams_perceptual_1995,mermelstein_distance_1976}. In this line of research, subjects are asked to rate the distance between multiple pairs of recorded instrument notes. The differences are used to fit to a timbre space in which all instruments are situated. The dimensions accounting for the greatest variance in this space are then correlated to acoustic descriptors of the stimuli. One particularly strong acoustical correlate is the spectral centroid, a measure that appears independently in many timbre studies \cite{mcadams_confirmatory_2002}. Spectral centroid expresses the weighted mean of the spectral frame normalized by frame energy---in a frame containing a single sinusoid, the spectral centroid will be its frequency. It is a strong acoustical correlate of perceived ``brightness'' \cite{peeters_timbre_2011}.

\subsection{Automatic Audio-to-Audio Distances}

Many important audio modeling settings rely upon a perceptually-coherent distance between two audio representations, both in generative and constrastive setttings. Generative audio losses can be viewed as divergences between audio representations. Thus, audio generation tasks use audio-to-audio distances: 
a) neural synthesis, in particular conditional spectrogram inversion \cite{Engel2017-sz,kalchbrenner2018efficient,mehri2016samplernn,oord_wavenet_2016};
b) audio source separation \cite{Hennequin2019-ii,Jansson2017-lc,Pariente2020-it,vincent_deep_2015};
c) denoising \cite{Germain2018-pp,Manocha2020-cn};
and
d) upsampling \cite{zhang_deep_2020}, not to mention audio self-supervised learning in general.

LeCun \cite{lecun2020-keynote} argues that most human and animal learning is self-supervised, whether generative or contrastive. Contrastive losses are based upon knowing that the representation distance between the target and a \emph{similar} prediction is lower than the representation distance between the target and a \emph{dissimilar} prediction. The contrastive approach involves fewer parametric assumptions than the generative one, and so we use it in our dissimilarity-ranking experiments (see Section~\ref{sec:evaluation-measures}).

\subsection{Learning Representations of Pitch}

Knowing that pitch is an important dimension for representation learning, there are several options for inducing pitch in learned representations: data, features, and loss functions.

Engel et al. \cite{Engel2020-tc} supplement their {\em data} with synthesized audio of a known pitch to aid in self-supervised pitch learning.
One may use explicit pitch-{\em features} from a pitch extractor, such as CREPE \cite{kim_crepe_2018} or SPICE \cite{DBLP:journals/taslp/GfellerFRSTV20} as input \cite{Hantrakul2019-cj, luong_investigating_2018} or conditioning information \cite{kalchbrenner2018efficient,oord_wavenet_2016,yasuda_investigation_2019}. Nonetheless, learning new data and incorporating better features is gated by the underlying loss function.

Audio {\em loss} construction typically involves some prior knowledge from digital signal processing, for example in their choice of time-frequency representation. What is surprising is that these well-motivated distances perform poorly in our simple pitch experiments. This negative result explains why methods using spectral losses nonetheless end up using pitch-tracker models for conditioning \cite{engel_ddsp_2020} or reconstruction \cite{Engel2020-tc}.
Alternately or in addition to DSP-motivated losses, knowledge from auxiliary models can be distilled using feature-matching losses (Section~\ref{sec:differentiable-distances}).

\section{Distances based on Spectral Representations}

Auditory neural networks commonly use spectral representations of sound. One can generate a (static) spectrum from a time-domain signal with the Fourier transform in Equation~\ref{eq:fft}:

\begin{tabular}{p{5cm}p{7cm}}
\begin{equation}
\label{eq:fft}
   X(\omega) = \int_{-\infty}^{+\infty}x(t)e^{-j\omega t} dt
   \end{equation}
&
   \begin{equation}
   \label{eq:stft}
X(\tau, \omega) = \int_{-\infty}^{\infty}x(t)w(t-\tau)e^{-j\omega t} dt
\end{equation}
\end{tabular}

where $x(t)$ is the signal, $t$ is time, $\omega$ is frequency and $j = \sqrt{-1}$ is the imaginary unit. The Fourier transform exchanges all time information for frequency information, which is not always desirable for analysis purposes. In order to keep both, the signal can be windowed into small segments or ``frames.'' Taking a Fourier transform of each frame results in a two dimensional function called the short-time Fourier transform (STFT), shown in Equation~\ref{eq:stft},
where $\tau$ is the time index and $w(t)$ is the windowing function. As with the Fourier transform, the STFT produces a complex-valued function. Its real-valued modulus is sufficient for many applications; this magnitude-only version is commonly called the ``spectrogram,'' and indicates the amplitude of frequency components over time. The spectrogram and its variants (e.g., the Mel spectrogram) are among the most popular audio representations for signal analysis.

\begin{figure}[t]
  \centering
  \begin{subfigure}{0.208\textwidth}
    \includegraphics{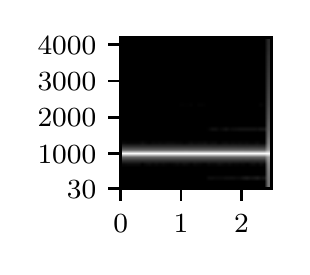} 
  \end{subfigure}
  \begin{subfigure}{0.15\textwidth}
    \includegraphics{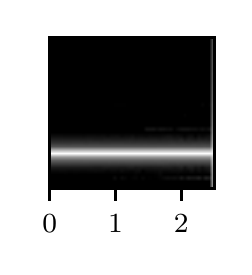} 
  \end{subfigure}
  \begin{subfigure}{0.15\textwidth}
    \includegraphics{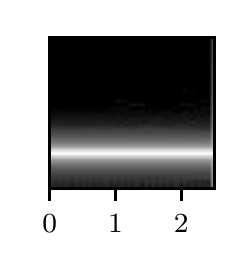} 
  \end{subfigure}
  \begin{subfigure}{0.15\textwidth}
    \includegraphics{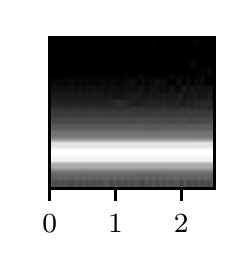} 
  \end{subfigure}
  \begin{subfigure}{0.15\textwidth}
    \includegraphics{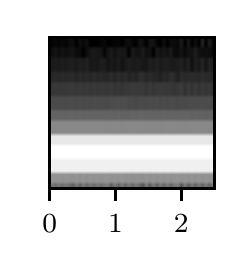} 
  \end{subfigure}
  \begin{subfigure}{0.15\textwidth}
    \includegraphics{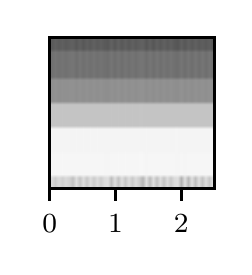} 
  \end{subfigure}

  \caption{Spectrograms for a 1000 Hz sine tone sampled at 44.1kHz, shown for various window sizes: 2048, 1024, 512, 256, 128, and 64 (left-to-right). $x$-axis is time in seconds and $y$-axis is Hz.}
  \label{fig:stft}
\end{figure}

All time-frequency representations suffer from the same fundamental trade-off, which is a direct consequence of the uncertainty principle \cite{gabor_theory_1947}:
the more that is known about a signal's location in time, the less is known about its frequency, and vice versa. Shorter STFT frames gain temporal accuracy, but cut off frequencies that require a longer time extent to fully oscillate (Figure~\ref{fig:stft}).
Window functions $w(t)$ are designed to fade frames in and out, and introduce their own spectral profile into the signal. The choice of window is a trade-off between reducing the spectral noise floor (lower ``side-lobes'') and gaining spectral precision (narrower ``main lobes'') \cite{harris_use_1978}. We use a Hann window throughout.

\subsection{Differentiable Audio Distances}
\label{sec:differentiable-distances}

We are interested in {\em differentiable} audio distances that are well-behaved and easy to optimize. Audio-to-audio distances are typically the $\ell_1$ or $\ell_2$ between representations of the target audio and the predicted audio. While the raw waveform can be used to this end \cite{kalchbrenner2018efficient,mehri2016samplernn,oord_wavenet_2016}, a more common approach is to use the $\ell_1$ or $\ell_2$ distance between the spectrogram or log-magnitude spectrogram \cite{NIPS2018_8118,Ping2018-wc,Wang2017-xj,Yamamoto2019-ol}.

A particular loss might have certain weaknesses in certain situations, such as gradient instability. Multiple losses may be mixed to work in tandem, and hopefully add up to more than the sum of their parts. The multi-scale spectrogram (MSS) captures difference in audio phenomena at multiple scales of time and frequency in an attempt to side-step the time-frequency trade-off discussed in the previous section. Interestingly, this roughly echoes a strategy used in the primary auditory cortex \cite{chi_multiresolution_2005}. As before, multi-scale losses are typically computed as $\ell_1$ or $\ell_2$ over linear- and log-spectrograms \cite{
    Bitton2020-wy,
    dhariwal2020jukebox,
    Engel2020-tc,
    engel_ddsp_2020,
    wang_neural_2019,
    Yamamoto2020-ir,
    Yang2020-sy}.

Alternately, one may use neural auditory distances. In the ``knowledge-distillation'' or ``teacher-student'' framework, one net trains the other \cite{DBLP:journals/corr/HintonVD15}.  A specific formulation of this approach, sometimes called ``perceptual loss,'' or ``feature-matching,'' considers the internal activations of parts or the whole of a neural net \cite{Engel2020-tc,kumar_melgan_2019,Ping2018-wc,zhang_unreasonable_2018}. Even specifically for pitch-learning, there are numerous examples of a feature-matching losses: a typical speech VQ-VAE is pitch-insensitive \cite{NIPS2017_7a98af17}, but one can add a pitch reconstruction loss \cite{DBLP:conf/icassp/GarbaceaOLLLVW19} or learn an additional pitch-only VQ-VAE \cite{DBLP:conf/interspeech/0006LLWCY20}.

\section{Experimental Design}
\label{sec:experimental-design}
Our synthetic benchmark suggest that losses based upon common spectral representations, as well as many off-the-shelf neural representations, have difficulties tracking pitch orientation. Our tool for this inquiry is a plain model of pure stationary sine waves: a well-behaved audio-to-audio loss should have high pitch-orientation accuracy, both on a fine scale as well as a coarse one. This approach is similar in spirit to \cite{2020arXiv200904709B}, which uses gradient-orientation as an evaluation criterion on a toy spheres task (albeit in the setting of adversarial learning). 

\subsection{Pure Sinuisoids}
\label{sec:the-problem}

Pure sine waves have three underlying factors of variation: amplitude ($A$), frequency ($\omega$), and initial phase ($\phi$):

\begin{equation} x_t(A, \omega, \phi) = A\cos(\omega t + \phi)
    \label{eq:sine-wave}
\end{equation}

Given a target sine wave, we seek an audio distance that permits a sine-wave model to generate a predicted sine wave with the same frequency and level as the target. 
Note that while linear amplitude and frequency are used to generate the signal, these quantities are more meaningfully described by their logarithmic equivalents \cite{stevens_scale_1937}. Thus, in our benchmark, sinusoids are generated with a uniform distribution in the \emph{pitch space} (or logarithmic distribution in frequency) between 30--4000Hz. The \emph{signal level} (hereafter, ``level'') is sampled from a normalized uniform decibel scale 0db through -25dB. We calculated normalized level of a sinusoid as \begin{equation} L_A = 25\log_{10}(A),
\end{equation} such that a maximum amplitude $A$ of 1 corresponds to a signal level $L_A$ of 0dB.
The initial phase $\phi$ of isolated sinusoids is not perceptible. We randomize it to remove a confounding variable.

\begin{figure}[ht]
\label{fig:both-distances}
    \centering
    \begin{subfigure}[b]{0.33\textwidth}
        \includegraphics[width=\linewidth]{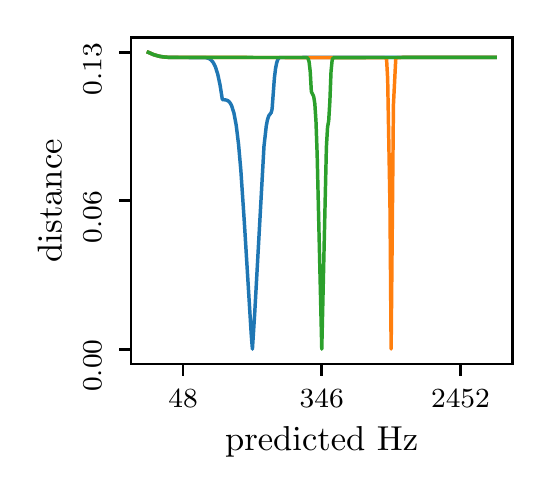}
	\caption{Spectrogram}
	\label{fig:distance-curve-spectrogram}
    \end{subfigure}
    \begin{subfigure}[b]{0.33\textwidth}
        \includegraphics[width=\linewidth]{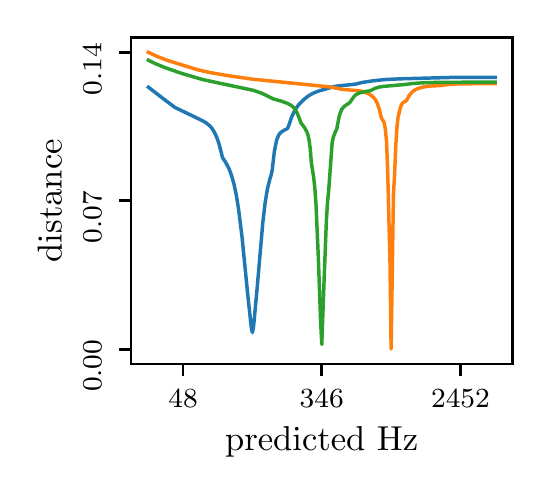}
	\caption{MSS}
	\label{fig:distance-curve-msstft}
    \end{subfigure}
  \caption{In the above figures, the three curves show distances from target pitches fixed at $130, 346$, and $922$ Hz. The target and prediction signal level are fixed at -12.5dB.
Figure~\ref{fig:distance-curve-spectrogram} shows the spectrogram-based distance between the target and prediction, as the predicted pitch $\omega$ varies.
Figure~\ref{fig:distance-curve-msstft} shows the MSS (multi-scale spectrogram) distance.
}
  \label{fig:distance-curve}
\end{figure}

\subsection{Tuning A Sine: Thought Experiment}

Imagine that an experimenter asks you to adjust the pitch of one sine tone oscillator to match another. If the pitches start far apart, you would have a confident sense of the direction to tune in, up or down, and have gradually less confidence as you approach the target pitch.

Figure~\ref{fig:distance-curve-spectrogram} shows the spectrogram-based distance, as predicted pitch sweeps across the spectrum. The observed pitch sensitivity is in many ways the opposite of a natural pitch sense: at great distances it has no sense of pitch orientation whatsoever (a vanishing gradient), and it locks-in with extreme confidence at fine pitch distances.
This result stands to reason, given the coordinate-for-coordinate comparison implicit in norm-based distance calculations---an effect that is only exaggerated in spectrograms with large frame size, and therefore fine spectral resolution (Figure~\ref{fig:stft}, left). The MSS-based distance (Figure~\ref{fig:distance-curve-msstft}) somewhat alleviates this problem, but still suffers from a sharp gradient near the target and a vanishing gradient in high predicted pitch extrema. Finally, our distances are measured on a log-linear frequency scale. However, the spectrogram operates on a linear scale, so lower target pitches have slightly wider regions with informative gradients. 

\subsection{Gradient-Sign (Dissimilarity) Ranking Accuracy}
\label{sec:evaluation-measures}

In a well-behaved distance, a target sinusoid should be closer to a predicted candidate sinusoid than to a {\em perturbed} candidate sinuisoid that is further away by construction. This argument motivates an evaluation using a dissimilarity-ranking accuracy \cite{pmlr-v2-agarwal07a,DBLP:conf/icml/OuyangG08}, i.e.\ whether the gradient is pointed in the correct direction.

We randomly sample a target sinusoid with level $A$ and pitch $\omega$, and a prediction sinusoid with level $A'$ and pitch $\omega'$ (where $A \ne A'$ and $\omega \ne \omega'$). We also construct a third perturbed sinuisoid. The perturbed signal level is the same as the prediction signal level. The perturbed signal pitch $\omega'' := \omega' + \varepsilon\cdot\sgn(\omega' - \omega)$ is, by construction, $\varepsilon$ further away from the target $\omega$ than the prediction signal pitch $\omega'$. From these signals we compute two distances: the distance from the target to the prediction, and the distance from the target to perturbation. In a well behaved distance, the prediction should be closer to the target than the perturbation is. This condition gives us a 0/1 error. Repeated and averaged over 1000 trials, high accuracies indicate that gradients usually point in the right direction. We reproduce the same process using perturbations in level with a fixed pitch, using $A'' := A' + \varepsilon\cdot\sgn(A' - A)$. Formally:
\begin{align}
    {\omega}\ \textrm{Acc} := \underset{\omega, A, \omega', A''}{\textrm{mean}} \mathbbold{1}\big (d(x(A, \omega), x(A', \omega')) < d(x(A, \omega), x(A', \omega'') \big)
\label{eq:pitch_sign_accuracy}\\
    {A}\ \textrm{Acc} := \underset{\omega, A, \omega', A''}{\textrm{mean}} \mathbbold{1}\big(d(x(A, \omega), x(A', \omega')) < d(x(A, \omega), x(A'', \omega')\big)
\label{eq:level_sign_accuracy}
\end{align}

In addition to numeric computations using explicit values of $\varepsilon$, a gradient may be solved and computed analytically using automatic differentiation. This is the equivalent of letting this value go to zero in Equations~\ref{eq:pitch_sign_accuracy} and~\ref{eq:level_sign_accuracy}. We define $\epsilon := \varepsilon \to 0$.

We make no assumptions about pitch- and level- distance, besides our choice of range (30Hz-4KHz, -25dB to 0dB) and that sampling occurs logarithmically with respect to $A$ and $\omega$ (Equation~\ref{eq:sine-wave}).

\section{Results}
\label{sec:results}

We present quantitative results in Table~\ref{tbl:results}. Each column indicates the average accuracy for the analytic, fine- and coarse-numeric conditions, with respect to changing pitch and changing level. The sign-accuracy of analytic gradients ($\epsilon \approx 0$) demonstrates whether the distance is capable of breaking out of very fine-grained local minima. While in practice gradient descent uses the analytic gradient, we also present numerically-computed gradients to characterize local minima at broader resolutions. The {\em fine} perturbation values (30 cents or 2dB) are meant to be plausibly detectable by a discerning human listener. The {\em coarse} perturbation values (600 cents or 10dB) are meant to be obvious to a human listener with normal hearing, and to give an indication of broad-scale local minima in the search space. 

For qualitative elucidation, we provide coarse heatmaps based on a fixed target pitch and level in Figure~\ref{fig:heatmaps-all}. Many distances are sensitive to the choice of target signal (Figures~\ref{fig:distance-curve-spectrogram} and \ref{fig:distance-curve-msstft}), so heatmaps based on a fixed target do not necessarily illustrate the entire search space. Supplementary Figures~\ref{fig:heatmaps-more1} and \ref{fig:heatmaps-more2} depict coarse heatmaps for other target pitches and levels.

\subsection{Spectral Audio Distances}

The \textbf{spectrogram} (the modulus of Eq.~\ref{eq:stft}) is a starting point for many kinds of spectral analysis. A spectrogram-based distance can roughly model pitch in the coarse condition (69.5\%), but otherwise the gradients score near chance (i.e., they are not pointed in any meaningful direction). Level-perturbed gradients score near random in both coarse and fine conditions. In the case of both pitch and level, the analytical gradient is in agreement with the fine perturbation conditions. Interestingly, the log-magnitude spectrogram has some level sensitivity, but is worse at pitch tracking than its linear equivalent.

\textbf{Mel spectrograms} are a popular variant of the spectrogram based on the Mel scale \cite{stevens_scale_1937}. By pooling frequency bins into progressively wider frequency spans, this representation roughly emulates the log-scaled nature of human frequency perception. As with the spectrogram, the Mel spectrogram-based distance performs near chance (if slightly above) in all conditions.
 
\begin{figure}[p]
    \centering
    \begin{subfigure}[b]{0.16\textwidth}
        \label{fig:heatmap-ideal}
        \includegraphics[width=\linewidth]{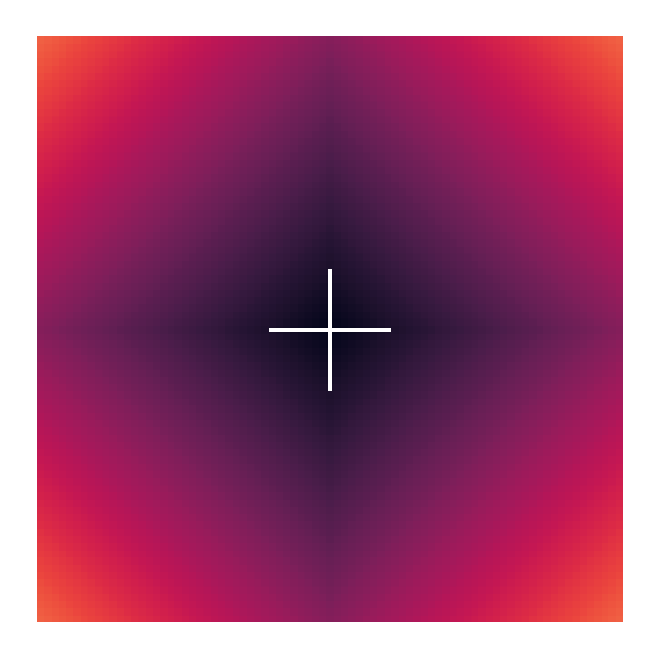}
        \caption{\centering Idealized $\ell_1$ \newline}
         \label{fig:heatmaps-ideal}
    \end{subfigure}
    \hfill
    \begin{subfigure}[b]{0.16\textwidth}
        \includegraphics[width=\linewidth]{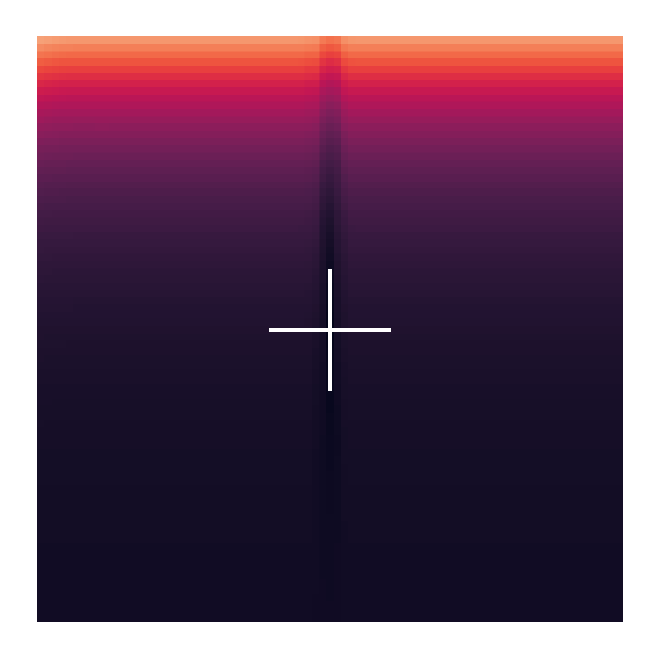}
        \caption{\centering Spectrogram \newline}
        \label{fig:heatmaps-stft}
    \end{subfigure}
    \hfill
    \begin{subfigure}[b]{0.16\textwidth}
        \includegraphics[width=\linewidth]{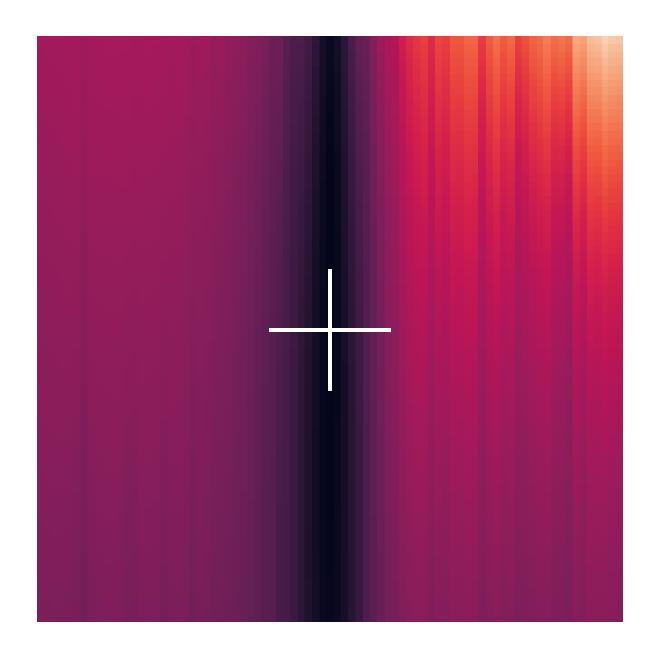}
        \caption{\centering log Spectrogram}
        \label{fig:heatmaps-logstft}
    \end{subfigure}
    \hfill
    \begin{subfigure}[b]{0.16\textwidth}
        \includegraphics[width=\linewidth]{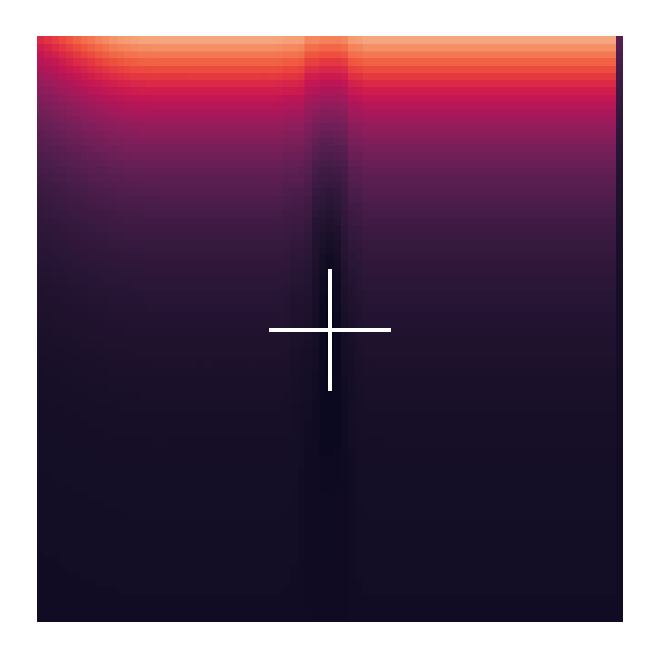}
        \caption{\centering Mel \newline}
        \label{fig:heatmaps-mel}
    \end{subfigure}
    \hfill
    \begin{subfigure}[b]{0.16\textwidth}
        \includegraphics[width=\linewidth]{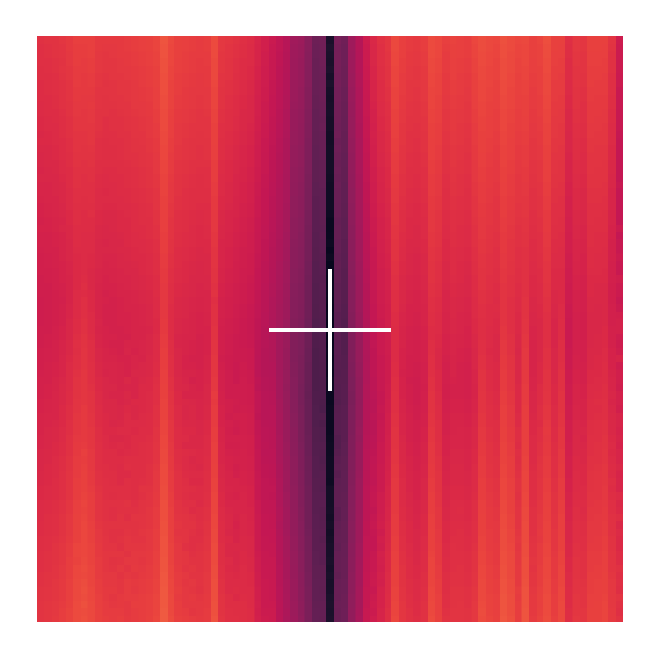}
        \caption{\centering MFCC \newline}
        \label{fig:heatmaps-mfcc}
    \end{subfigure}
    \hfill
    \begin{subfigure}[b]{0.16\textwidth}
        \includegraphics[width=\linewidth]{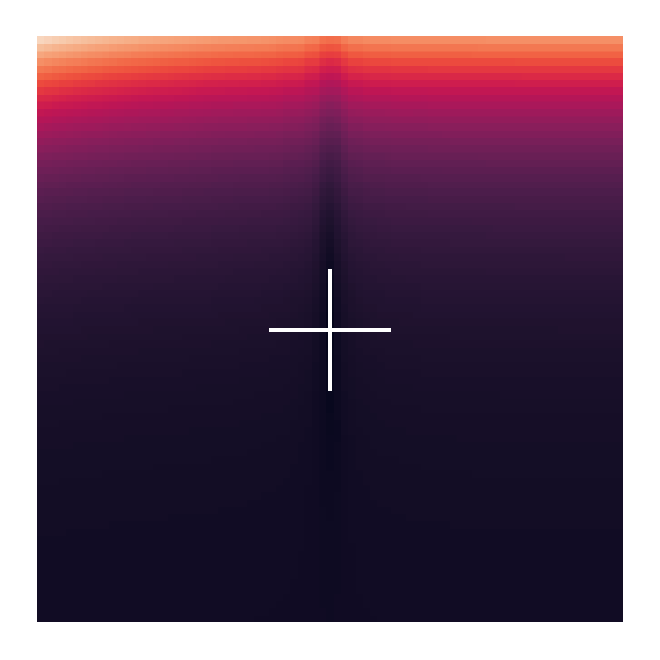}
        \caption{\centering MSS \newline}
        \label{fig:heatmaps-msstft}
    \end{subfigure}

    \begin{subfigure}[b]{0.16\textwidth}
        \includegraphics[width=\linewidth]{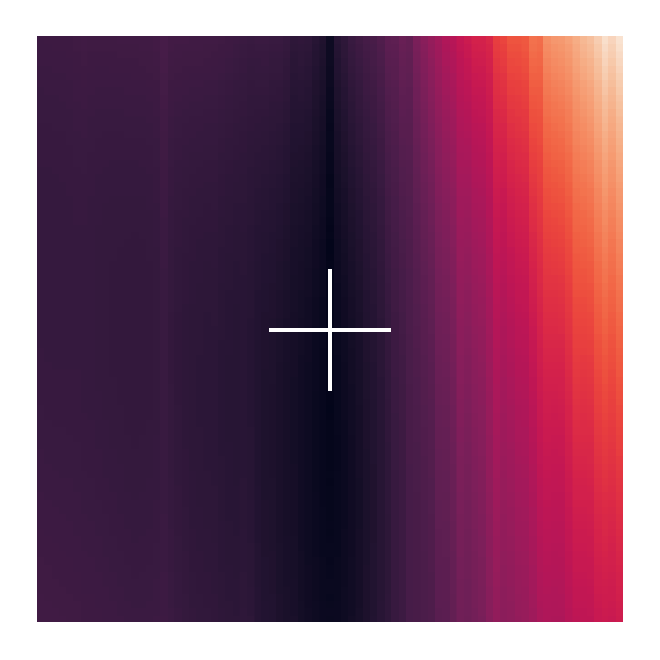}
        \caption{\centering log MSS \newline}
        \label{fig:heatmaps-logmsstft}
    \end{subfigure}
    \hfill
    \begin{subfigure}[b]{0.16\textwidth}
        \includegraphics[width=\linewidth]{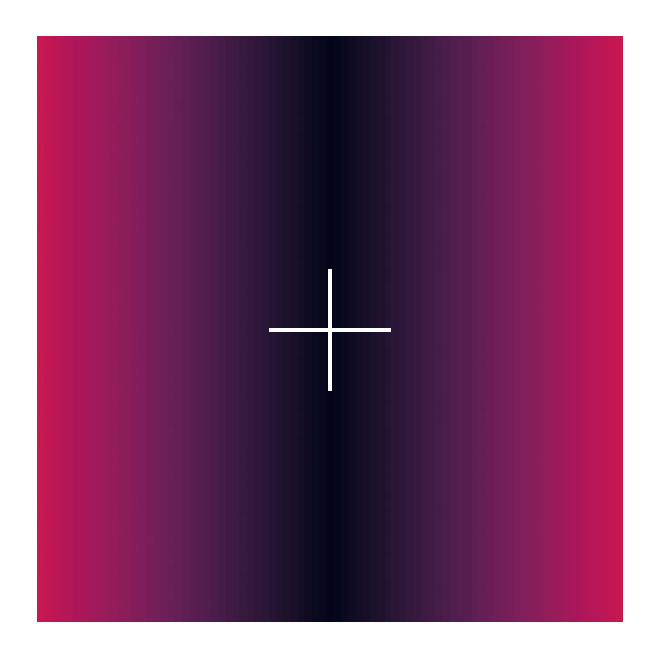}
        \caption{\centering Spectral Centroid}
        \label{fig:heatmaps-sc}
    \end{subfigure}
    \hfill
    \begin{subfigure}[b]{0.16\textwidth}
        \includegraphics[width=\linewidth]{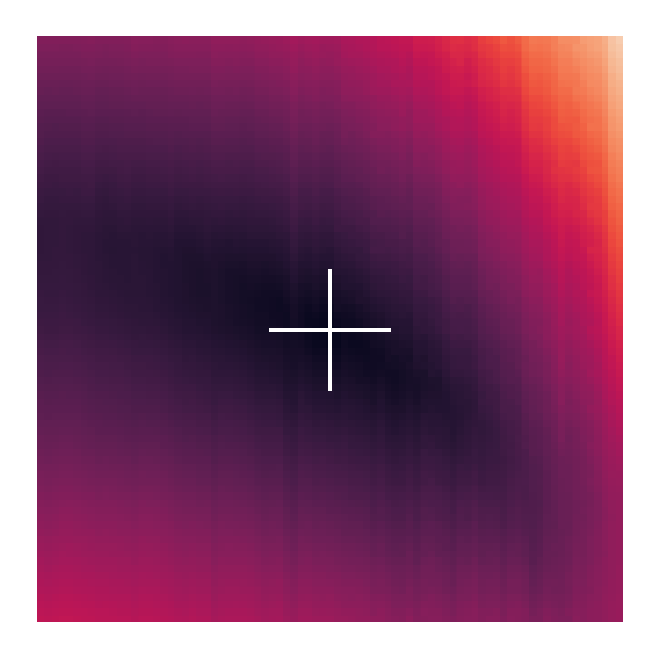}
        \caption{\centering NSynth Wavenet}
        \label{fig:heatmaps-wavenet}
    \end{subfigure}
    \hfill
    \begin{subfigure}[b]{0.16\textwidth}
        \includegraphics[width=\linewidth]{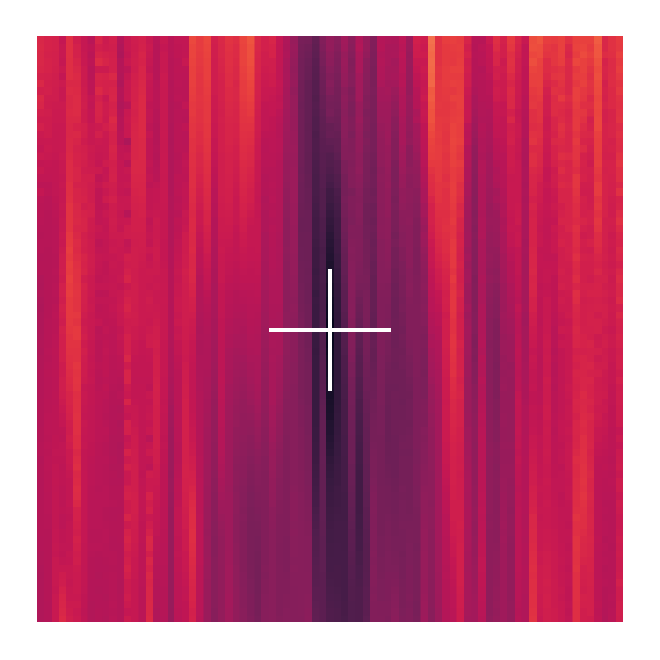}
        \caption{\centering VGGish \newline }
        \label{fig:heatmaps-vggish}
    \end{subfigure}
    \hfill
    \begin{subfigure}[b]{0.16\textwidth}
        \includegraphics[width=\linewidth]{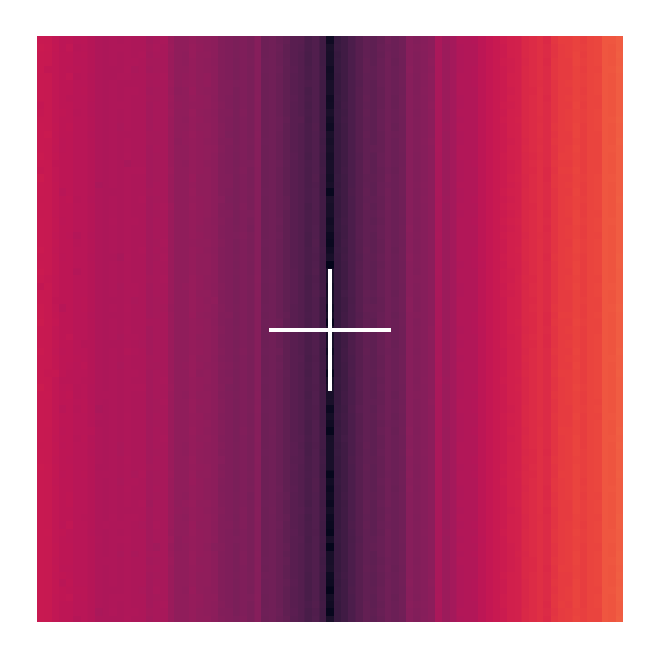}
        \caption{\centering OpenL3 \newline }
        \label{fig:heatmaps-openl3}
    \end{subfigure}
    \hfill
    \begin{subfigure}[b]{0.16\textwidth}
        \includegraphics[width=\linewidth]{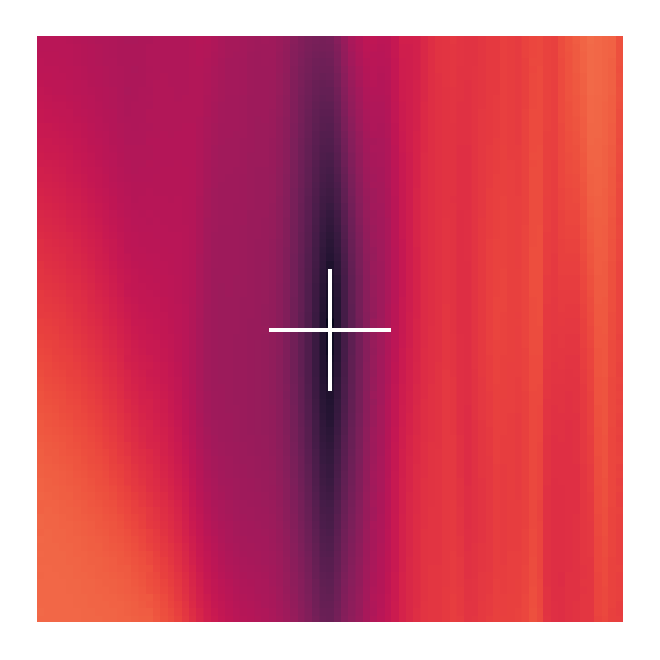}
        \caption{\centering wav2vec2 LV-60}
        \label{fig:heatmaps-wav2vec2}
    \end{subfigure}
  \caption{Measured distance between the target and prediction. Target pitch and level are fixed in the center of the search space ($346$ Hz and -12.5dB, respectively). The prediction pitch ($x$-axis) spans 30--4000Hz and the prediction level ($y$-axis) spans -25 dB--0 dB. The resolution of these heatmaps is 106 cents and 0.31 dB per cell. Figures~\ref{fig:heatmaps-more1} and \ref{fig:heatmaps-more2} show heatmaps for different targets.
}
  \label{fig:heatmaps-all}
\end{figure}

\textbf{Mel-frequency cepstral coefficients} (MFCCs) are a popular feature for speech and music applications based on cepstral analysis \cite{noll_cepstrum_1967}. In \cite{Wang2017-xj}, the authors argue that MFCCs are somewhat pitch invariant. Fittingly, the MFCC representation struggles with pitch, having an accuracy slightly worse than the spectrogram, though it has the best level score among the spectral representations we tested. The MFCC analytic gradient computed on level is entirely at odds with the fine perturbation results, scoring nearly at chance. This phenomenon is reflected in the extra fine-grained vertical streaks in the MFCC heatmap (Figure~\ref{fig:heatmaps-mfcc}). At a broader resolution the curves look smooth; at super-fine resolutions they are rocky.

\begin{table}[p]
\small
\centering

\begin{tabular}{r|ccc|ccc}

Candidate & $\omega$ Acc & $\omega$ Acc & $\omega$ Acc & $A$ Acc & $A$ Acc & $A$ Acc \\
Distance & $\pm \epsilon$ & $\pm$ 30 & $\pm$ 600 & $\pm \epsilon$  & $\pm$ 2 & $\pm$ 10 \\
& & cents & cents &  & dB & dB \\

\hline

\multicolumn{1}{l|}{Spectral representations}  \\


Spectrogram 
& 0.617 & 0.574 & 0.695 & \bf 0.535 & \bf 0.518 & \bf 0.514 \\

log(Spectrogram) 
& 0.548 & \bf 0.541 & 0.679 & 0.607 & 0.577 & 0.615 \\

Mel 
& \bf 0.511 & \bf 0.479 & \bf 0.580 & 0.564 & \bf 0.544 & 0.551 \\

MFCC 
& \bf 0.532 & 0.603 & 0.648 & 0.593 & \em 0.931 & \em 0.999 \\

MSS 
& 0.771 & \em 0.905 & \em 0.978 & 0.550 & \bf 0.530 & \bf 0.531 \\

log MSS 
& \bf
0.532 & 0.665 & \em 0.879 & 0.719 & 0.730 & 0.783 \\

$\log_2$(Spectral Centroid) 
& \bf 0.515 & \em 1.000 & \em 1.000 & \bf 0.460 & \bf 0.503 & \bf 0.518 \\

\hline

\multicolumn{1}{l|}{Neural audio representations}  \\

nsynth wavenet
& & 0.588 & \em 0.862 &  & \em 0.873 & \em 0.938 \\

vggish 
& & \bf 0.536 & 0.652 &  & 0.595 & 0.636 \\

openl3 
& & 0.594 & \em 0.989 &  & \bf 0.507 & \bf 0.480 \\

wav2vec 2.0 Large (LV-60), no fine-tuning 
& 
& 0.727 & \em 0.831 & 
& 0.682 & 0.738 \\

%
%

\hline

\end{tabular}

\caption{Accuracy of gradient orientation, at various resolutions, over 1000 trials. The worst case 95\% CI is $\pm 0.032$.
For spectral representations, resolution $\epsilon \to 0$ is the analytically-computed gradient, followed by fine-grained and coarse-grained numerically-computed gradient-orientation accuracy.
Bold indicates accuracy $\le 55\%$, near random gradient orientation at this resolution. Italics indicates accuracy $\ge 80\%$, which is likely to escape local minima using stochasticity.}

\label{tbl:results}
\end{table}

\begin{figure}[p]
    \centering
    \begin{subfigure}[b]{0.28\textwidth}
        \includegraphics[width=\linewidth]{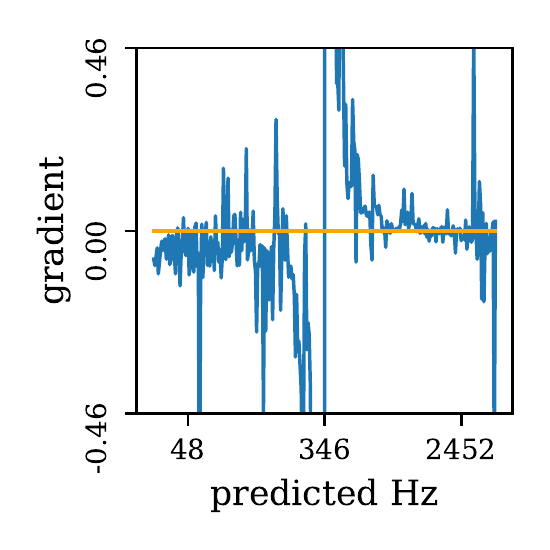}
	\caption{\centering Analytic MSS gradient}
	\label{fig:analytic-gradient-msstft}
    \end{subfigure}
    \begin{subfigure}[b]{0.28\textwidth}
        \includegraphics[width=\linewidth]{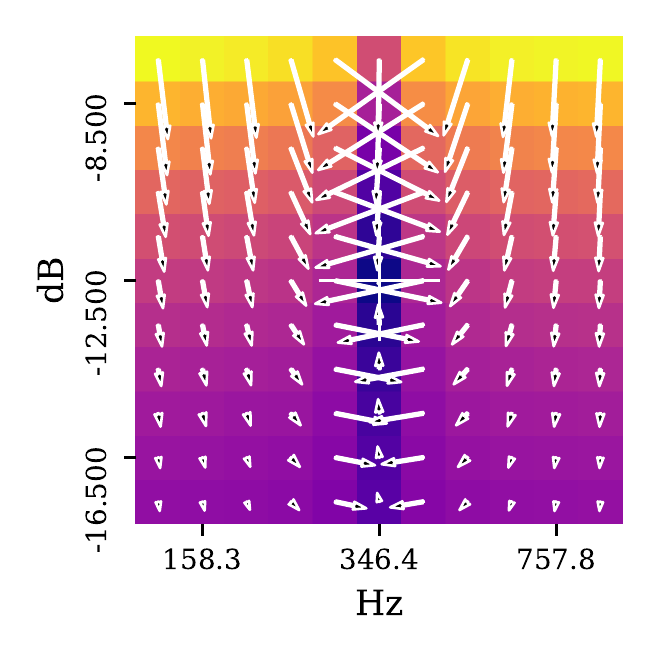}
	\caption{\centering Numeric MSS gradient}
	\label{fig:numeric-gradient-msstft}
    \end{subfigure}
  \caption{\ref{fig:analytic-gradient-msstft} shows the analytic gradient of MSS, with levels fixed at -12.5dB and target pitch $\omega = 346$ Hz. Figure~\ref{fig:numeric-gradient-msstft} shows a 2.5x zoomed in heatmap of the MSS distance. As in the heamaps of Figure~\ref{fig:heatmaps-all}, the target pitch and level are the center ($\omega=346$ Hz, $A = -12.5dB$). However, the resolution of 340 cents and 1 dB per cell. Arrows illustrate the numeric gradients at this resolution.
}
  \label{fig:gradients}
\end{figure}

\textbf{Multi-scale spectral loss} (MSS) compares spectrograms at multiple scales---in our work, we use the six window sizes shown in Figure~\ref{fig:stft} for MSS. In effect, this mitigates the poor behavior of the spectrogram-based pitch distance (Figure 2). Indeed, the pitch-perturbed gradient is correctly oriented 77\% of the time. Still the gradient is not entirely smooth (Figure~\ref{fig:analytic-gradient-msstft}). MSS has poor analytic level gradients (55\%). The heatmap in Figure~\ref{fig:numeric-gradient-msstft} shows the level-gradient always drives the level down and away from the target signal, even when the prediction level is lower than than the target. Only when pitch is locked-in are the level gradients correctly oriented. Hence, MSS level accuracies are 55\%. The log MSS outperforms the log-spectrogram, improving mainly on level but not pitch.

The log-scaled \textbf{spectral centroid} recovers the pitch precisely on pure sinusoids. As per its normalization factor, it is insensitive to level. However, its analytic gradients are highly chaotic for both level and pitch, scoring at chance.

\subsection{Neural Audio Distances}

Large-scale neural audio models are far more expressive than traditional spectral representations. Neural models have evinced uncanny ability on other zero-shot problem domains, such as uncovering analogical reasoning when trained to predict a missing word \cite{pennington2014glove}. We test a variety of models under our evaluation methodology, using their final layer as an audio representation.

\textbf{NSynth Wavenet} is trained on musical notes from the NSynth dataset \cite{Engel2017-sz}. It has a strong coarse level accuracy (93\%) and coarse pitch accuracy (86\%). At finer pitch distances the pitch accuracy suffers (59\%). This is perhaps because the dataset comprises single notes, each spaced one semitone (100 cents) apart.

\textbf{VGGish} \cite{hershey_cnn_2017} and \textbf{OpenL3} \cite{cramer_look_2019} are large-scale nets trained on a variety of audio samples taken from internet video, the latter incorporating multi-modal information from the video feed. Interestingly, both are insensitive to level, scoring just slightly above chance. OpenL3 is highly pitch accurate at a broad scale (99\%), but scores near random (59\%) at a fine scale.

We auditioned many pre-trained models of \textbf{wav2vec 2.0} \cite{baevski2020wav2vec}, a large-scale neural net trained completely on speech; we achieved the best results using their large (LV-60) no fine-tuning model. This model has good coarse-pitch accuracy (83\%), despite having never listened to music.

Overall, we were surprised to find some deficiencies of these models on our benchmark, which speaks to the difficulty of extracting a general-purpose low-dimensional representation of audio. Poor results of these representations is not meant to impugn whether pitch can be extracted from these representations using one or more layers. Fine-tuning a large-scale neural model on a simple task will likely succeed, even in a one-shot learning scenario \cite{zhang_unreasonable_2018}.

\section{Discussion}

Pitch and level are both natural dimensions to consider when comparing sounds. To a listener with typical hearing, making small adjustments based on pitch or level is as simple as tuning a guitar, or turning down the radio. So why is it that common spectral-based loss functions have difficulty with our benchmark? While some distances fare relatively well on a strictly pitch-based perturbation, almost all do poorly when introducing any perturbations of level. A further mystery still, why is the search space apparently bumpy at very fine resolutions?

In any norm-based distance, one makes the implicit assumption of orthogonality; i.e., that the only important difference between two vectors lies in the difference of \emph{corresponding} values. In the case of the spectrogram, the distance asks the question ``do I need more of frequency \emph{x}, or less?''. But these differences contain no explicit spectral orientation---in other words, they have no sense of their spectral neighbors, and by consequence can't answer a question that is simple for humans: ``is pitch \emph{x} higher or lower than pitch \emph{y}?''.

MSS appears to mitigate this problem by taking advantage of the spectral smearing of smaller window sizes (see Figure~\ref{fig:stft}). However, its analytic gradient remains unimpressive in the pitch condition, and nearly random in the level condition.

Many of the analytic gradients perform poorly. Even the 77\% accurate MSS gradient is not smooth (Figure~\ref{fig:numeric-gradient-msstft}). We do not completely understand the nature of this phenomenon, however in our investigations we have found a few patterns: unwindowed STFT distances result in a much bumpier gradient, and Hann windowing smooths the search space. The use of multiple analysis scales further smooths the search space, however these distances still contain many very fine-grained local minima.

Finally, pitch-recovery is an intrinsically difficult regression problem. Sinuisoidal activations can compactly construct functions with infinite VC-dimension \cite{Gaynier1995-wl}, and backprop through a periodic function is difficult to train.

\section{Does my non-music model really need robust pitch sensitivity?}
\label{sec:arguments-for-pitch-sensitivity}

A pessimistic perspective might be that if a particular task is almost solved, it is not useful to pursue broader machine understanding. If so, what could motivate the speech, NLP, and other machine listening communities
to include broad-scale pitch perception in their training regimes?

Task-specific approaches can produce one-trick ponies with unexpected failure modes. Sturm \cite{sturm_simple_2014} describes Clever Hans, a horse that could ostensibly do basic arithmetic, fooling even a zoologist. In fact, Clever Hans had no mathematical understanding at all, and instead relied upon subconscious visual cues from his interrogator. Bello \cite{bello-youtube} describes an analogous failure-mode: a model that predicted the tango genre exceptionally well, but only because all the tango recordings in its corpus were old and characteristically band-limited. This phenomenon is similar to certain patch-based face recognition approaches, where a face is recognized when presented with an adversarial example where the eyes and mouth are switched. Even a few pixel change can trick a neural network \cite{Su2019-ru,42503}.

There is a growing understanding that uncovering unusual failure modes is not butterfly hunting, but indeed is deeply connected with generalization. Adversarial examples occur in nature \cite{45818}. Gilmer et al.~\cite{Gilmer2018-gy} demonstrate a tight correlation between vision performance on a toy adversarial sphere dataset with generalization on natural tasks. Spurious correlation \cite{DBLP:journals/corr/abs-2010-15775} and under-specification  \cite{DBLP:journals/corr/abs-2011-03395} also lead to poor generalization.

Evaluation should not only quantify performance on a specific task, but aid the improvement of systems in general \cite{sturm_simple_2014}. Training on broad open-domain data, like the recently released FSD50K corpus \cite{DBLP:journals/corr/abs-2010-00475}, exposes the model to the full gamut of natural phenomena. Sufficiently diverse synthesized data might encompass real-world data as a special case \cite{DBLP:conf/iros/TobinFRSZA17}.
Multi-task learning demonstrates that sample complexity reduces, training accelerates, and generalization improves by including data that surfaces more information than is present in the target task's data \cite{collobert2011natural,ruder_overview_2017}.

Incorporating more training data may lead to a better result, but there might be a ceiling to what data alone can bring to the table \cite{Bisk2020}.
In NLP, the nuances of a question like ``is an orange more like a baseball or more like a banana?'' cannot be fully comprehended by a model trained purely on text, even the entire web. Learning the size of objects using pure text analysis requires significant gymnastics \cite{elazar2019large}; vision demonstrates physical size more easily.
This school of thought suggests that listening to all the audio on the web can only learn so much; true general purpose learning can only happen through grounding in an experiential notion of sound, i.e., one that is perceptually, and perhaps even cognitively, based. By extension, this implies incorporating more modalities than just hearing \cite{cramer_look_2019}.

More controversial still is the argument that passively consuming the internet's entirety of media data in all modalities (text, vision, and audio) provides insufficient context, and the broader contextual scopes of embodiment and social interaction are needed for deeper understanding \cite{Bisk2020}.

\section{Conclusion}

Learning is only as good as the underlying optimization criterion. Our synthetic benchmark exposes a surprising gap between automatic audio-to-audio distances and human perception. In contrast to the human experience of sound, spectrally-based audio representations---as well as several off-the-shelf neural representations---have difficulties with basic pitch orientation. When signal level is included as a dimension of variability, gradient orientation accuracy decreases. The confounding effects of level are perplexing, and further work is necessary to identify the cause.

Our negative results demonstrate that even for a low-dimensional problem (i.e., pitch and level), distances based upon these audio representations are perceptually unintuitive and bad for learning. Their search spaces have local minima at fine and coarse resolutions. These issues are surely compounded in higher-dimensional problems, and raise the spectre that they cannot model as-yet-unknown underlying factors of variation. The question remains: what is necessary to induce general-purpose audio representations similar to human auditory perception?

\section*{Broader Impact}

We present this work as an invitation to researchers to scrutinize auditory representations at large, and to hold loss functions to a higher standard, one closer to what is perceptually salient. We do not expect any negative outcomes from this work.

Models of auditory perception are implicitly based on prototypical notion of ``healthy'' hearing. We acknowledge that hearing ability varies widely from listener to listener, and that this line of research may be considered exclusionary from the perspective of the hearing impaired. To this extent, research in this direction may leverage data that are not wholly representative of the population. We should be cognizant of this bias and---indeed---it should be possible for machines to model different people's pitch perception.

\begin{ack}

The authors would like to thank the following readers for their feedback:
Alex Nisnevich,
Alexandre Passos,
Edouard Oyallon,
Felipe Tobar,
Harri Taylor,
Jesse Engel,
Lamtharn (Hanoi) Hantrakul,
Nicolas Pinto,
Oleg Polosin,
Pranay Manocha,
Richard Socher,
Richard Zhang,
as well as the workshop reviewers and chairs.

\end{ack}

\bibliographystyle{plain}
\bibliography{../jptpaperpile,../refsmore,../max}

\appendix

\section{Supplementary Material}

\subsection{Ideal Models of Pitch Distance}
\label{sec:pitch-ortho}

There are at least two orthogonal factors when in comes to a representation of pure pitch distance:
\begin{enumerate}
    \item Distance determined by the ratio of the fundamental frequencies of the notes in question. For example, the ratio of 660:440Hz is equivalent to the ratio 750:500Hz, and therefore their pitch distances are the same. However, this rule is complicated by the fact that human hearing is not equally sensitive to pitch throughout the spectrum. We typically hear melodic information in the range from ~30Hz---4KHz \cite{Oxenham2012-mg,stevens_scale_1937}.
    \item Notes separated by one or more octaves (frequency ratio of powers of two, e.g.\ A440Hz and A880Hz) share a ``pitch class'' and are equivalent in this pitch dimension. This phenomenon leads to helical representations of pitch, which curls up and around once an octave \cite{lostanlen_learning_2020,Shepard1982-jt}.
\end{enumerate}

To the knowledge of the authors, the relative magnitude of these two dimensions in the perception of pitch distance is an open question. For simplicity, we model only the first dimension. Under this assumption, the ideal perceived pitch distance is strictly monotonic for fixed level \cite{stevens_scale_1937}:
\begin{align}
    d(x(A, \omega), x(A', \omega')) < d(x(A, \omega), x(A', \omega''))\ &\textrm{if}\ \omega < \omega' < \omega''\ \textrm{or}\ \omega > \omega' > \omega''
    \label{eqn:ordinal1}\\
    d(x(A, \omega), x(A', \omega')) < d(x(A, \omega), x(A'', \omega'))\ &\textrm{if}\ A < A' < A''\ \textrm{or}\ A > A' > A''
    \label{eqn:ordinal2}
\end{align}
which lead to Equations~\ref{eq:pitch_sign_accuracy} and \ref{eq:level_sign_accuracy}.

\subsection{Hyperparameters}

All spectral distances are implemented in PyTorch using a sampling rate of 44.1KHz and Hann windowing. Other windowing methods available in {\tt scipy} performed equivalently or worse. Although hyperparameter selection on evaluation data is against the spirit of zero-shot learning, we tuned hyperparameters to concisely demonstrate upper-bounds on performance. These are the hyperparameters used in our results section:

\begin{table}[htb]
\centering

\begin{tabular}{r|l}

Spectrogram & $\ell_1$ distance, $|\textrm{Spectrogram}(x)|$, nfft=2048, \\
                & frame\_overlap=0.75 \\
\hline

log(Spectrogram) & $\ell_2$ distance, $\log(|\textrm{Spectrogram}(x)| + $1e-4$)$, nfft=2048, \\
        & frame\_overlap=0.75 \\
\hline

Mel & $\ell_1$ distance, nfft=1024, frame\_overlap=0.5, nmels=1024,\\
                              & fmin=30, fmax=4000 \\
\hline

MFCC & $\ell_1$ distance, nfft=1024, frame\_overlap=0.5, nmels=128, \\
                               & nmfcc=128, norm=None, fmin=30, fmax=4000 \\
\hline

MSSTFT and log MSSTFT & Computed as Spectrogram and log(Spectrogram) respectively, \\
           & with nffts=[2048, 1024, 512, 256, 128, 64] \\
\hline

Spectral Centroid & $\ell1$ distance, $\log_2$ spectral centroid, \\
             & with nfft=2048, frame\_overlap=0.75, power=1.0\\

\end{tabular}

\caption{Hyperparameters used in spectral distances.}

\label{tbl:hyperparameters}
\end{table}

\begin{figure}[bhtp]
    \centering
    \begin{subfigure}[b]{0.16\textwidth}

        \includegraphics[width=\linewidth]{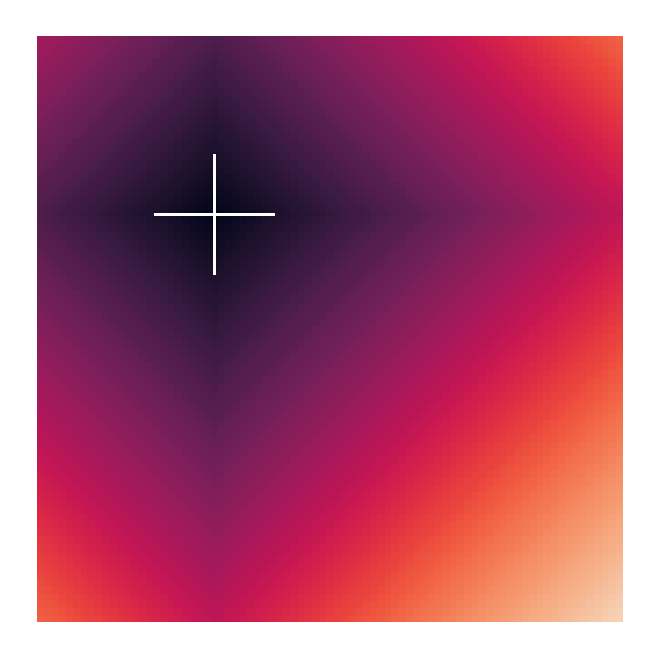}
        \caption{\centering Idealized $\ell_1$ \newline}

    \end{subfigure}
    \hfill
    \begin{subfigure}[b]{0.16\textwidth}
        \includegraphics[width=\linewidth]{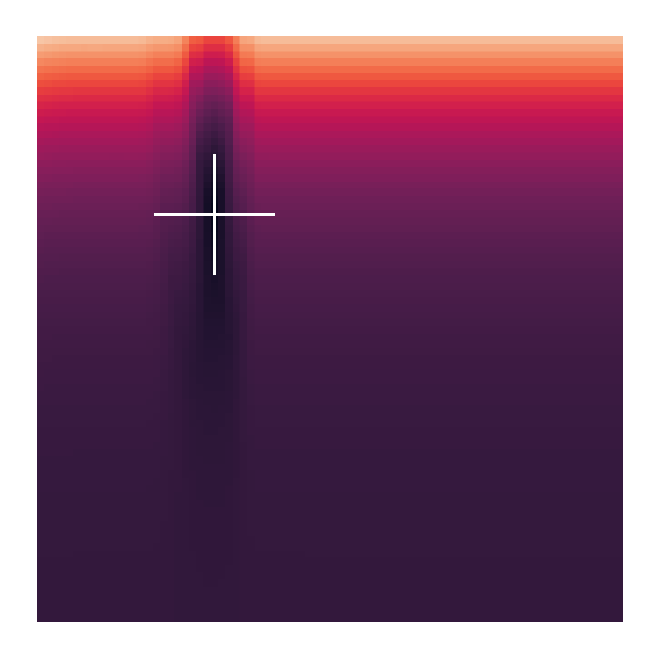}
        \caption{\centering Spectrogram \newline}

    \end{subfigure}
    \hfill
    \begin{subfigure}[b]{0.16\textwidth}
        \includegraphics[width=\linewidth]{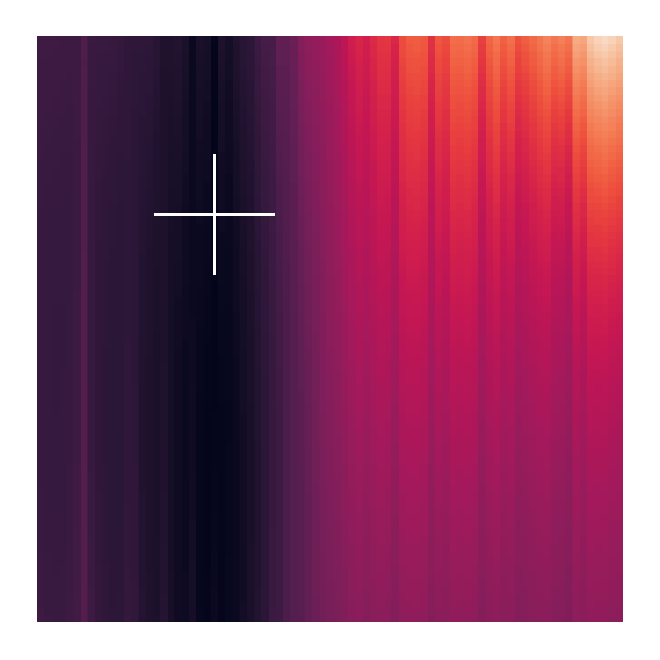}
        \caption{\centering log Spectrogram}

    \end{subfigure}
    \hfill
    \begin{subfigure}[b]{0.16\textwidth}
        \includegraphics[width=\linewidth]{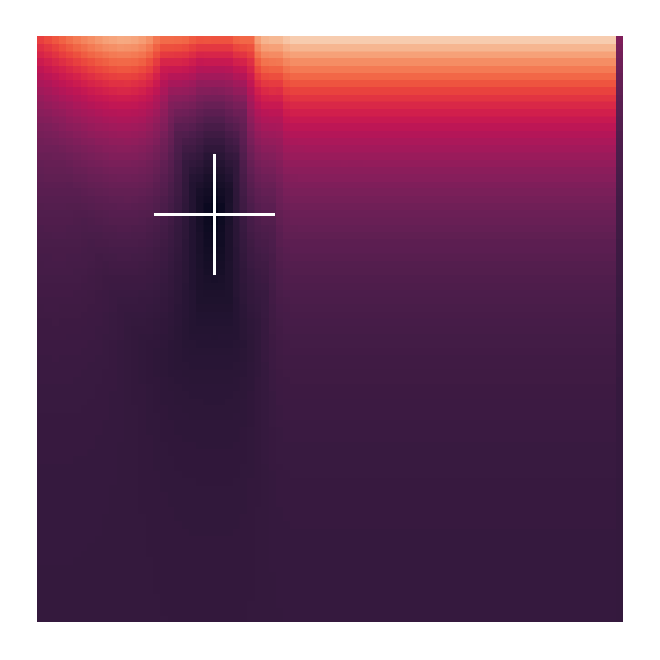}
        \caption{\centering Mel \newline}

    \end{subfigure}
    \hfill
    \begin{subfigure}[b]{0.16\textwidth}
        \includegraphics[width=\linewidth]{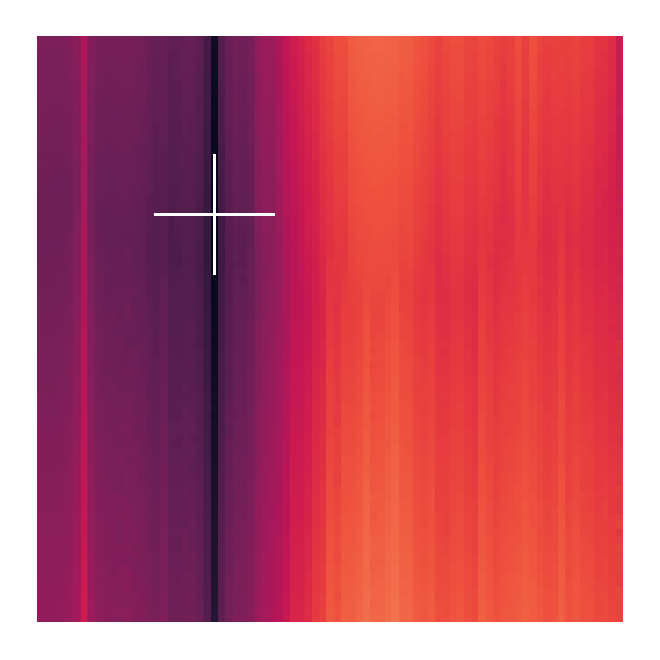}
        \caption{\centering MFCC \newline}

    \end{subfigure}
    \hfill
    \begin{subfigure}[b]{0.16\textwidth}
        \includegraphics[width=\linewidth]{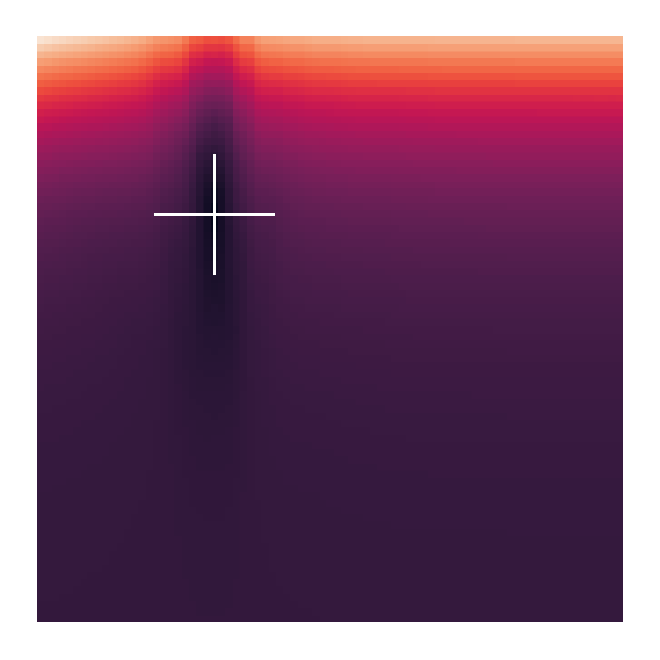}
        \caption{\centering MSS \newline}

    \end{subfigure}

    \begin{subfigure}[b]{0.16\textwidth}
        \includegraphics[width=\linewidth]{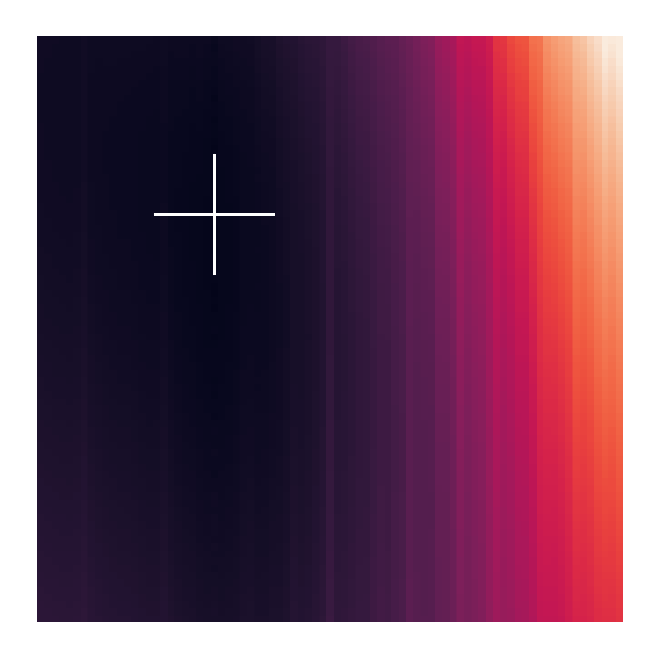}
        \caption{\centering log MSS \newline}

    \end{subfigure}
    \hfill
    \begin{subfigure}[b]{0.16\textwidth}
        \includegraphics[width=\linewidth]{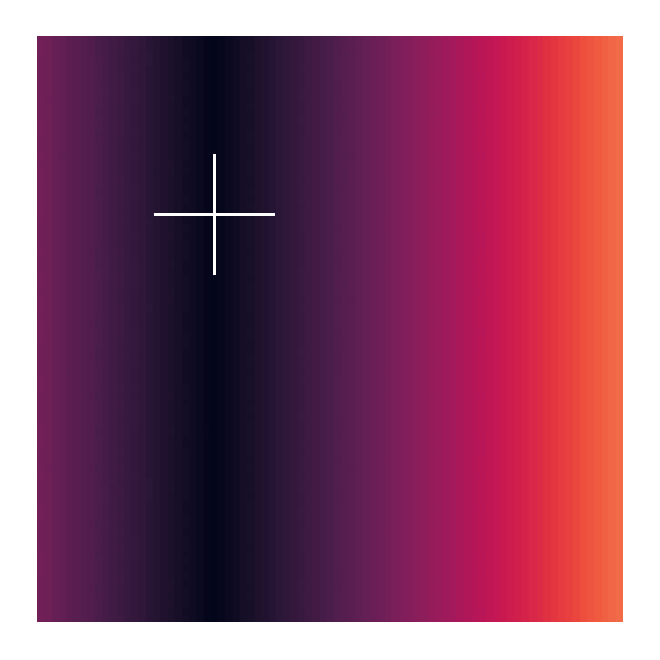}
        \caption{\centering Spectral Centroid}

    \end{subfigure}
    \hfill
    \begin{subfigure}[b]{0.16\textwidth}
        \includegraphics[width=\linewidth]{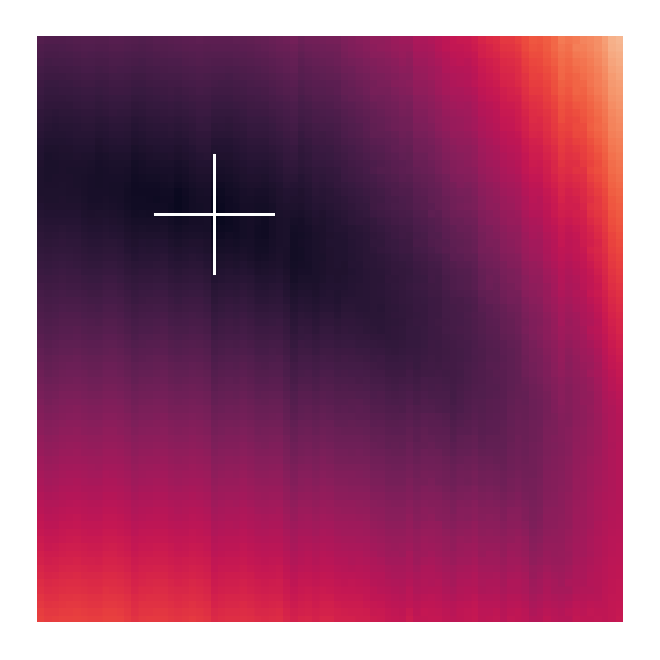}
        \caption{\centering NSynth Wavenet}

    \end{subfigure}
    \hfill
    \begin{subfigure}[b]{0.16\textwidth}
        \includegraphics[width=\linewidth]{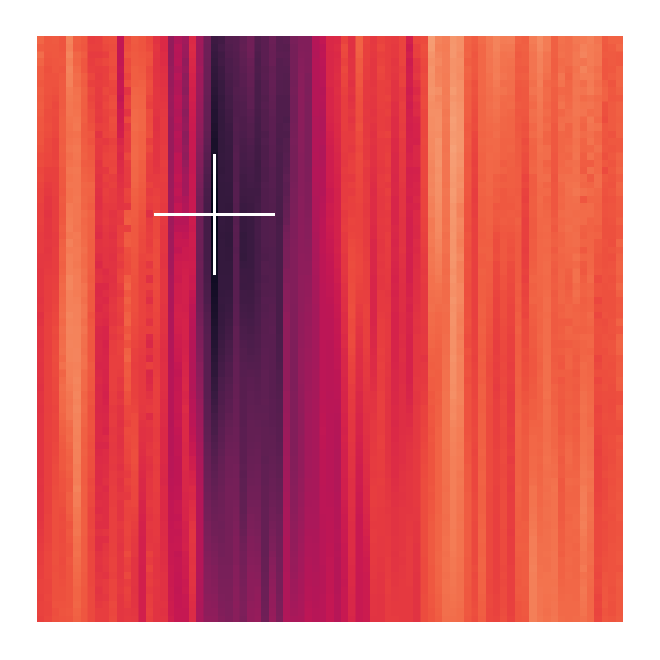}
        \caption{\centering VGGish \newline }

    \end{subfigure}
    \hfill
    \begin{subfigure}[b]{0.16\textwidth}
        \includegraphics[width=\linewidth]{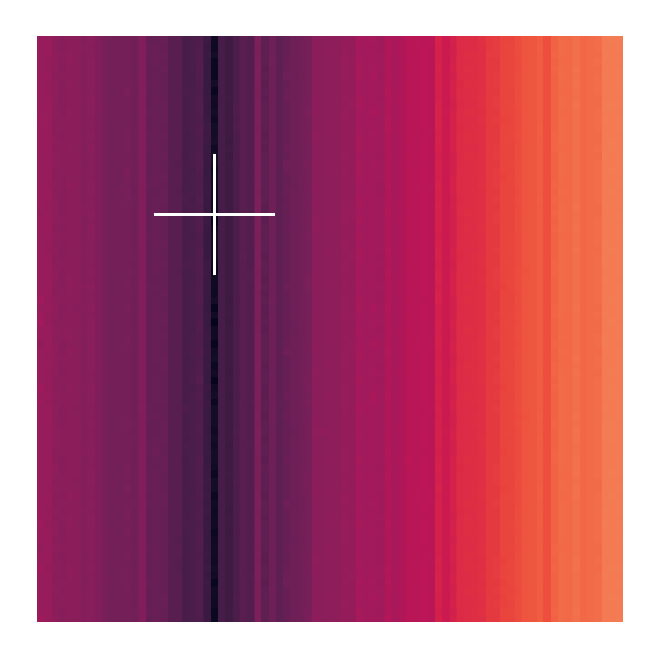}
        \caption{\centering OpenL3 \newline }

    \end{subfigure}
    \hfill
    \begin{subfigure}[b]{0.16\textwidth}
        \includegraphics[width=\linewidth]{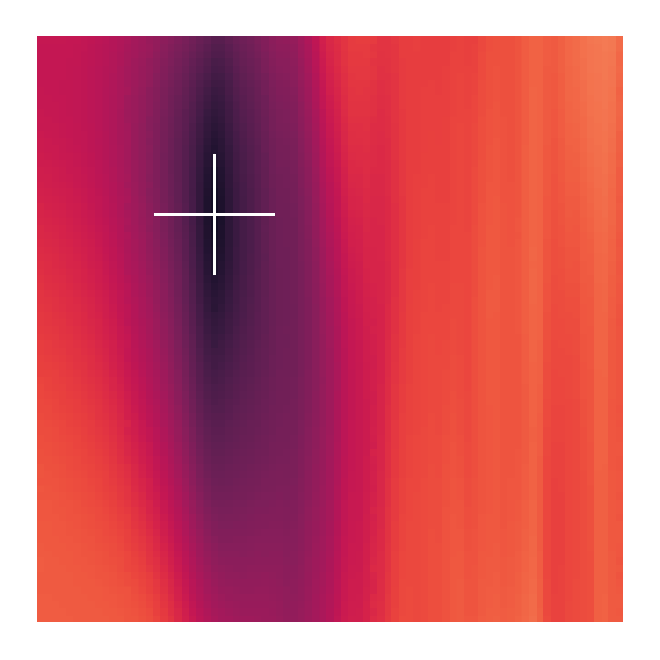}
        \caption{\centering wav2vec2 LV-60}

    \end{subfigure}
  \caption{As Figure~\ref{fig:heatmaps-all}, but with target pitch and level fixed at $130$ Hz and -7.5dB, respectively.}
  \label{fig:heatmaps-more1}
\end{figure}

\begin{figure}[ht]
    \centering
    \begin{subfigure}[b]{0.16\textwidth}

        \includegraphics[width=\linewidth]{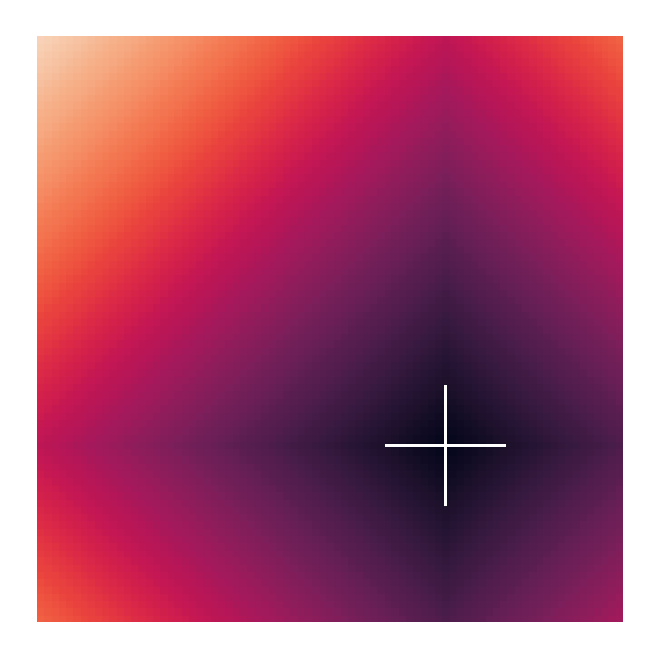}
        \caption{\centering Idealized $\ell_1$ \newline}

    \end{subfigure}
    \hfill
    \begin{subfigure}[b]{0.16\textwidth}
        \includegraphics[width=\linewidth]{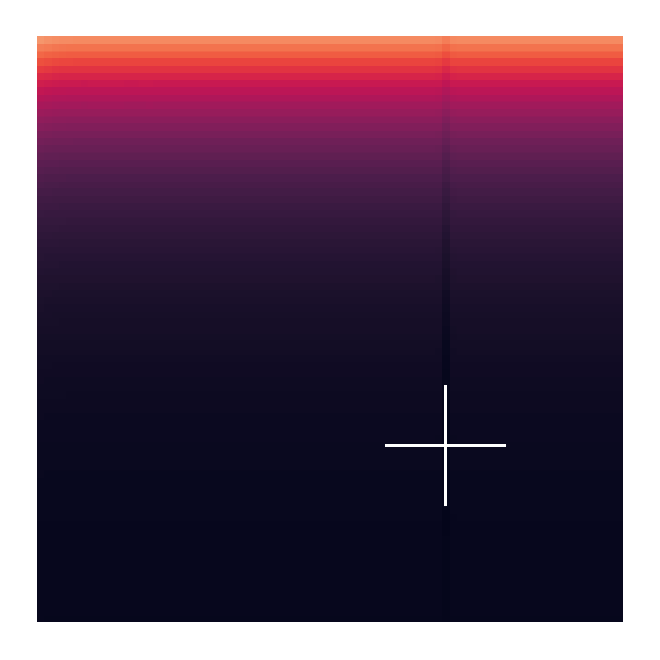}
        \caption{\centering Spectrogram \newline}

    \end{subfigure}
    \hfill
    \begin{subfigure}[b]{0.16\textwidth}
        \includegraphics[width=\linewidth]{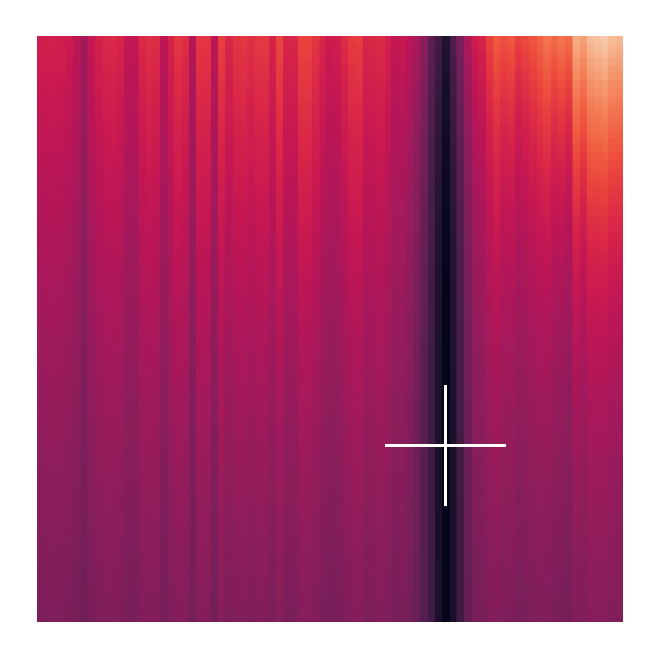}
        \caption{\centering log Spectrogram}

    \end{subfigure}
    \hfill
    \begin{subfigure}[b]{0.16\textwidth}
        \includegraphics[width=\linewidth]{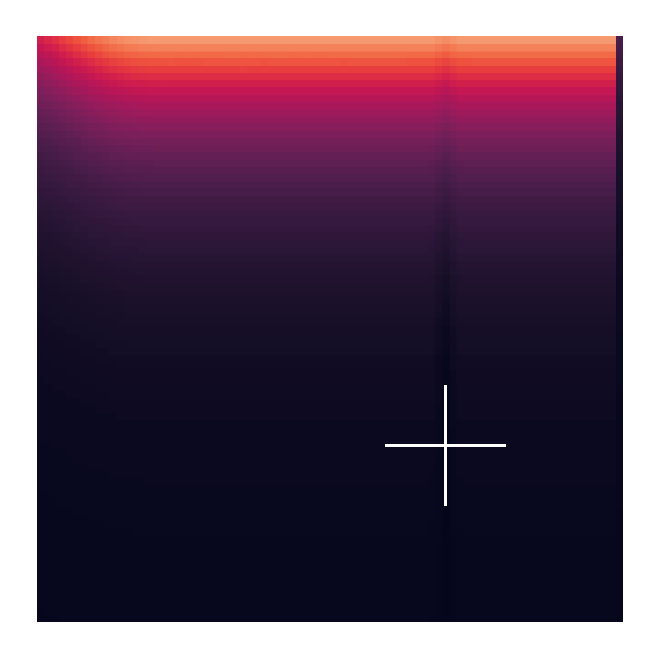}
        \caption{\centering Mel \newline}

    \end{subfigure}
    \hfill
    \begin{subfigure}[b]{0.16\textwidth}
        \includegraphics[width=\linewidth]{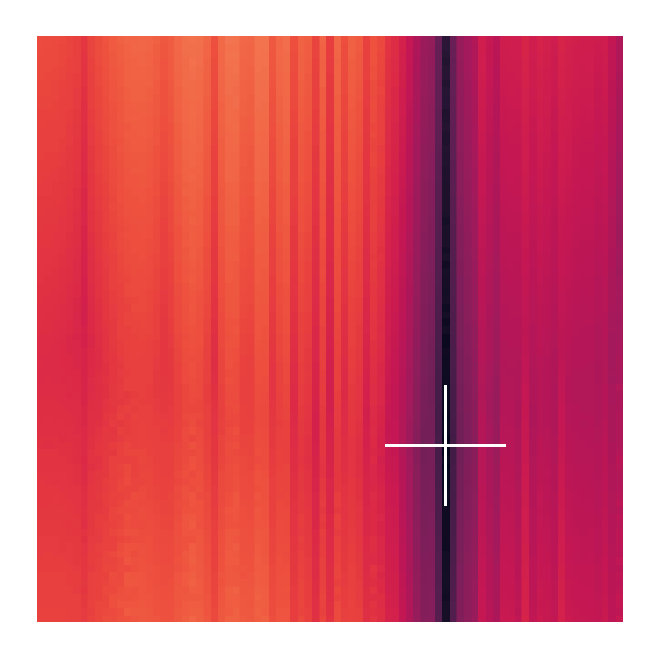}
        \caption{\centering MFCC \newline}

    \end{subfigure}
    \hfill
    \begin{subfigure}[b]{0.16\textwidth}
        \includegraphics[width=\linewidth]{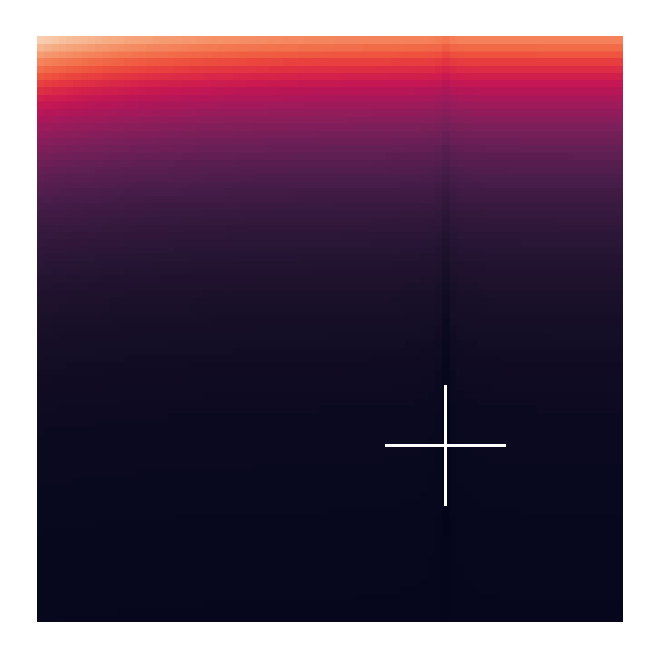}
        \caption{\centering MSS \newline}

    \end{subfigure}

    \begin{subfigure}[b]{0.16\textwidth}
        \includegraphics[width=\linewidth]{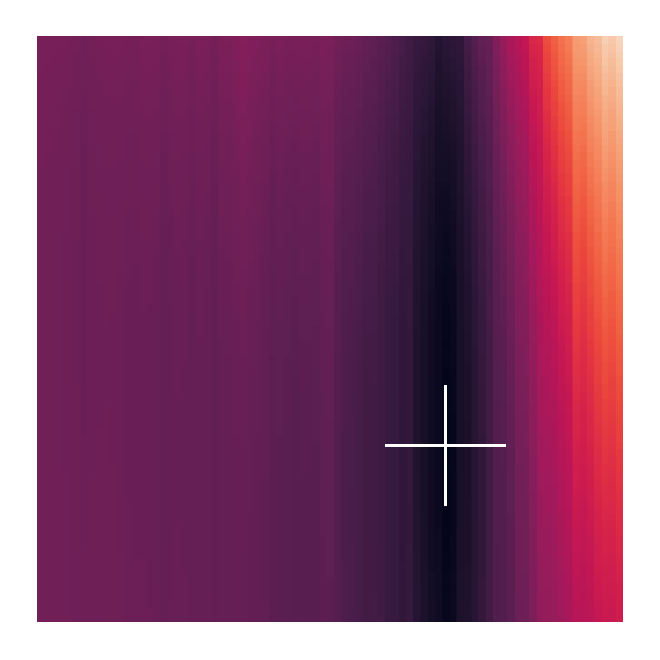}
        \caption{\centering log MSS \newline}

    \end{subfigure}
    \hfill
    \begin{subfigure}[b]{0.16\textwidth}
        \includegraphics[width=\linewidth]{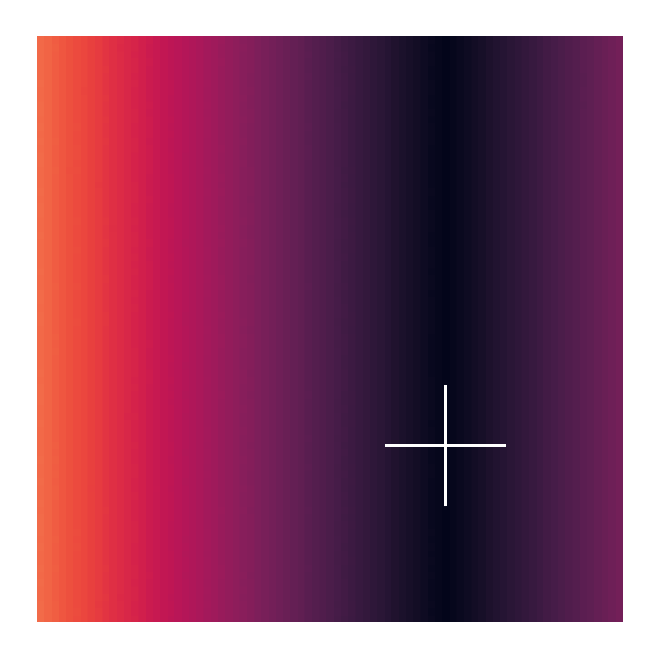}
        \caption{\centering Spectral Centroid}

    \end{subfigure}
    \hfill
    \begin{subfigure}[b]{0.16\textwidth}
        \includegraphics[width=\linewidth]{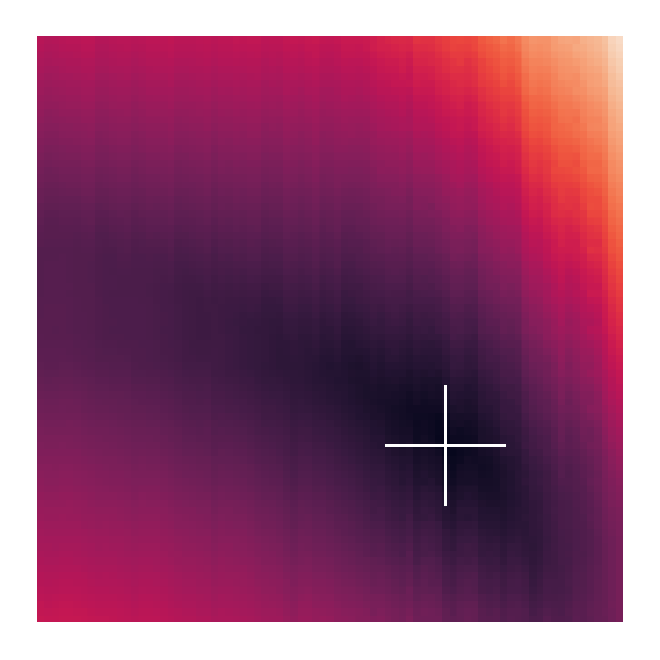}
        \caption{\centering NSynth Wavenet}

    \end{subfigure}
    \hfill
    \begin{subfigure}[b]{0.16\textwidth}
        \includegraphics[width=\linewidth]{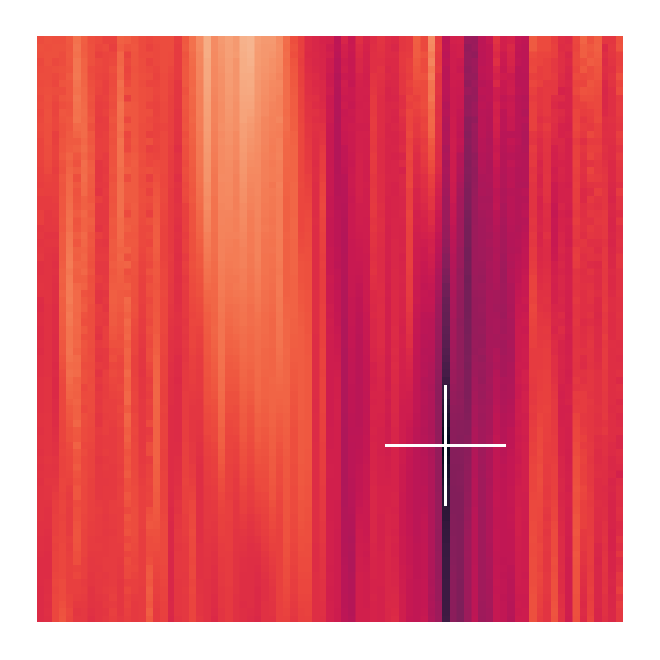}
        \caption{\centering VGGish \newline }

    \end{subfigure}
    \hfill
    \begin{subfigure}[b]{0.16\textwidth}
        \includegraphics[width=\linewidth]{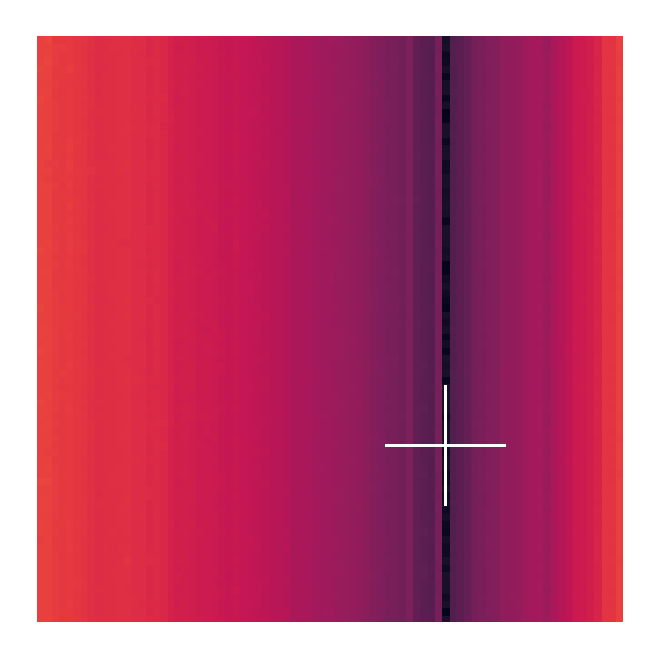}
        \caption{\centering OpenL3 \newline }

    \end{subfigure}
    \hfill
    \begin{subfigure}[b]{0.16\textwidth}
        \includegraphics[width=\linewidth]{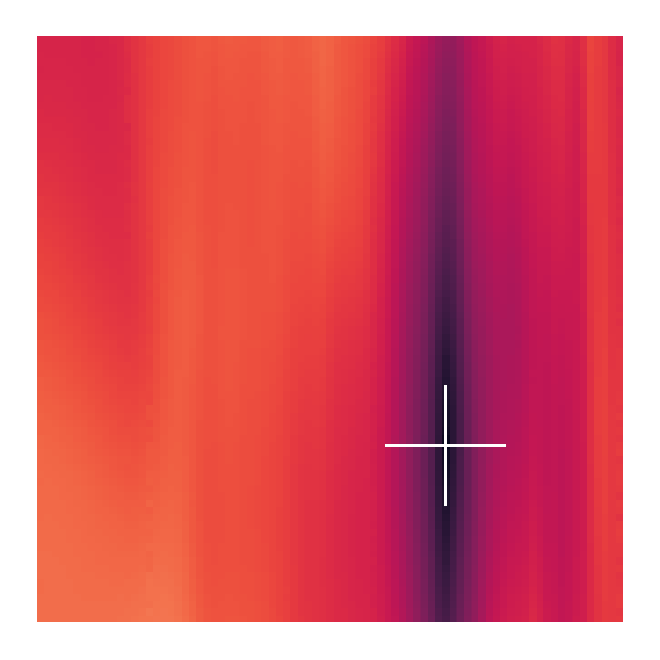}
        \caption{\centering wav2vec2 LV-60}

    \end{subfigure}
  \caption{As Figure~\ref{fig:heatmaps-all}, but with target pitch and level fixed at $922$ Hz and -17.5dB, respectively.}
  \label{fig:heatmaps-more2}
\end{figure}

\end{document}